\documentclass[aps,prd,twocolumn,showpacs,floats,floatfix,letterpaper,nofootinbib,superscriptaddress,notitlepage]{revtex4-1}
\pdfoutput=1
\usepackage{hyperref}
\usepackage{amssymb,amsmath,latexsym,mathrsfs}
\usepackage{graphicx,subfigure}
\usepackage{epsfig}
\usepackage{varioref,xr-hyper}
\usepackage{color}
\usepackage{multirow}
\usepackage{array}
\hypersetup{
     colorlinks   = true,
     citecolor    = cyan,
     linkcolor = magenta
     }
     
\begin{document}

\title{Light propagation in Swiss cheese models of random close-packed Szekeres structures: Effects of anisotropy and comparisons with perturbative results}
\author{S. M. Koksbang}
\email{koksbang@phys.au.dk}
\affiliation{Department of Physics and Astronomy, Aarhus University, 8000 Aarhus C, Denmark}


\begin{abstract}
Light propagation in two Swiss cheese models based on anisotropic Szekeres structures is studied and compared with light propagation in Swiss cheese models based on the Szekeres models' underlying Lemaitre-Tolman-Bondi models. The study shows that the anisotropy of the Szekeres models has only a small effect on quantities such as redshift-distance relations, projected shear and expansion rate along individual light rays.
\newline\indent
The average angular diameter distance to the last scattering surface is computed for each model. Contrary to earlier studies, the results obtained here are (mostly) in agreement with perturbative results. In particular, a small negative shift, $\delta D_A:=\frac{D_A-D_{A,bg}}{D_{A,bg}}$, in the angular diameter distance is obtained upon line-of-sight averaging in three of the four models. The results are, however, not statistically significant. In the fourth model, there is a small positive shift which has an especially small statistical significance.  The line-of-sight averaged inverse magnification at $z = 1100$ is consistent with $1$ to a high level of confidence for all models, indicating that the area of the surface corresponding to $z = 1100$ is close to that of the background.
\end{abstract}
\maketitle

\section{introduction}
It is standard to interpret cosmological observations simply with the spatially homogeneous and isotropic Friedmann-Lemaitre-Robertson-Walker (FLRW) models. Since the Universe is not exactly spatially homogeneous and isotropic, it is not {\em a priori} clear that this is a valid approach. However, it was argued in \cite{Linder1} that FLRW light propagation relations should be approximately valid for average observations in spacetimes exhibiting compensation of inhomogeneities along light rays. Such considerations were studied more thoroughly in \cite{syksy_av1,syksy_av2}, with the conclusion being that average observations should be well described by the volume-averaged dynamics and geometry in (spatially) statistically homogeneous and isotropic spacetimes with structures evolving slowly compared to the time it takes a light ray to pass through them, {\em if} light samples spacetime fairly
\footnote{A light ray is said to sample spacetime fairly if the averages of relevant quantities such as density, expansion rate etc. along the ray correspond well with the volume-averages of these quantities, with statistical deviations permitted.}
. Under these requirements, and to the extent that cosmic backreaction (see {\em e.g.} \cite{bc1,bc2,bc3,bc4}) and a possible local inhomogeneity can be neglected, it is then valid to use FLRW models to interpret observations. However, the requirement of tracing spacetime fairly is not necessarily fulfilled for observations involving point-like sources such as supernovae (see {\em e.g.} \cite{misinterp,Linder_angle}). The requirement {\em is} on the other hand expected to be fulfilled for observations related to sources with large angular extent such as the CMB \cite{Linder_angle}.
\newline\indent
As pointed out in {\em e.g.} \cite{bolejko_DA}, analyses of CMB observations depend crucially on the angular diameter distance to the last scattering surface, $D_{A,ls}$. It is therefore important to know if there are small systematic shifts in the redshift-distance relation in inhomogeneous spacetimes fulfilling the above listed requirements, compared to the relation obtained using the volume-averages of such spacetimes (which correspond to FLRW models when backreaction can be neglected). Studies involving both perturbation theory \cite{do_we_care,Kaiser_Peacock}, N-body simulations \cite{Nbody_average,Nbody_LTB_average} and exact, inhomogeneous solutions to Einstein's equations ({\em e.g.} \cite{syksyCMB,bolejko_DA,randomize,Ishak,tetris,tetris2,Nbody_LTB_average,ltb_effect_hubble}) imply that small shifts in the average value of $D_A$ due to inhomogeneities indeed occur. However, as discussed in {\em e.g.} \cite{do_we_care,Kaiser_Peacock}, this shift in $D_{A,ls}$ is already included in standard analyses of the CMB based on perturbation theory. This conclusion is only valid to the extent that a small average shift in $D_{A,ls}$ actually is described adequately by perturbation theory. It is in this respect important to acknowledge that perturbation theory is only an approximation scheme and that subtle effects important for precision cosmology may be beyond its reach. This makes it important to test results obtained with perturbation theory to make sure that they are actually true for spacetimes exhibiting, especially nonlinear, structure formation. One tool for performing such tests is Swiss cheese models based on the spherically symmetric Lemaitre-Tolman-Bondi (LTB) \cite{LTB1,LTB2,LTB3} structures glued together in areas where they have reduced to FLRW ``backgrounds" representing the volume-averages of the given Swiss cheese models. These (toy-)models give a description of spacetimes with energy-density fluctuations on well-defined average FLRW backgrounds and can thus be considered as describing ``exact perturbations" on FLRW backgrounds. This makes Swiss cheese models based on LTB structures excellent for testing perturbative results.
\newline\indent
By using Swiss cheese models based on LTB structures, it has been found that there are no large shifts in the average redshift-distance relation except as selection effects arising when only special light rays in a given spacetime are considered (see {\em e.g.} \cite{randomize} vs. \cite{bias1,bias2}). A noticeable exception is \cite{tardis} where a $30\%$ shift in the average redshift-distance relation compared to that computed using volume-averaged quantities was found at $z = 100$. However, it is not entirely clear to what extent this large shift is related to the occurrence of surface layers in the studied model.
\newline\newline
Results based on Swiss cheese models with LTB structures {\em should} be in line with perturbation theory; it has been shown with direct comparisons that perturbation theory based on (semi-)nonlinear LTB density fields and the corresponding exact velocity fields reproduces exact light propagation in LTB models very well \cite{dig_selv,dig_selv2} (see also \cite{randomize,LTB_in_NG,LTB_in_NG2,LTB_in_NG3,bright_side} for the relation between LTB models and perturbation theory). An important factor for the impressive reproductions of light propagation in LTB models with perturbation theory in \cite{dig_selv,dig_selv2} is the spherical symmetry of the LTB models; the reproduction is inhibited for anisotropic Szekeres structures because the method employed for obtaining velocity fields leads to artificial non-vanishing peculiar velocity fields outside non-spherically symmetric structures. That is, the method leads to non-vanishing peculiar velocity fields in regions where the exact anisotropic Szekeres models have reduced to FLRW models and hence should not give rise to peculiar velocity fields. This is unfortunate since real structures are not exactly spherically symmetric, making it important to know if spherical symmetry is essential for the agreement between Swiss cheese and perturbative results
\footnote{During the final stages of preparing this paper, a paper (\cite{Mexico}) relating Szekeres models to perturbation theory was published. The work presented there will be valuable for studies, especially theoretical ones, of how Szekeres models deviate from perturbation theory.}. It is in this respect notable that anisotropic Szekeres models exhibit structure formation that deviates significantly from both that of the underlying LTB models and from predictions of perturbation theory \cite{bolejko_struct1,bolejko_struct2, ishak_struct_vsLTB, ishak_struct_vsLin1, ishak_struct_vsLin2}. In addition, it is very clear from figure 7a in \cite{dig_selv2} that light propagation is significantly affected by anisotropy; in that figure, it is seen that an initially radial light ray following an exact geodesic of an anisotropic Szekeres model moves through a significantly different density field than a light ray traced simply by using the Born approximation. Such a result is not possible for light rays in LTB models since the spherical symmetry dictates that an initially radial light ray remains so.
\newline\newline
The above considerations imply the importance of studying whether the anisotropy of quasispherical Szekeres structures affects average observations of {\em e.g.} the angular diameter distance to the surface of last scattering. However, while there are ample studies involving light propagation in LTB models, only few studies involve observations in Swiss cheese models based on anisotropic Szekeres structures. In fact, only a single study of average observations in Swiss cheese models based on Szekeres models seems to exist, namely \cite{Ishak}. The results from that paper indicate a small average shift in the angular diameter distance compared to the background value. The study in \cite{Ishak} uses an ensemble average which is typical of Swiss cheese studies with an important exception being \cite{syksyCMB} where line-of-sight averages are used. As shown in \cite{angle_vs_ensemble} (see {\em e.g.} also \cite{angle_vs_ensemble_early,do_we_care}), second-order perturbation theory predicts that ensemble averages lead to a positive shift in the angular diameter distance while line-of-sight averages lead to a negative shift (with the shift defined as $\delta D_A :=\frac{D_A-D_{A,bg}}{D_{A,bg}}$ such that a negative shift corresponds to image magnification, and a positive to demagnification). In partial contradiction to this, positive shifts are obtained in both \cite{syksyCMB} and \cite{Ishak}, where the shift is, however, not statistically significant in the former. Furthermore, it has been shown that the ensemble average of the magnification $\mu:=\frac{D_{A,bg}^2}{D_A^2}$ is equal to 1 while its inverse averages to 1 upon line-of-sight averages \cite{weinberg,angle_vs_ensemble_early,angle_vs_ensemble, Kaiser_Peacock}. These results hold at least to second order in perturbation theory, insofar that perturbations to the areas of constant-redshift surfaces are negligible (which is generally assumed to be the case). However, the study in \cite{syksyCMB} finds a low probability of an average $\mu^{-1}$ equal to $1$ while the erratum to \cite{Ishak} describes a good agreement with an average value of $\mu^{-1}$ of 1.  In summary, the studies of average light propagation in Swiss cheese models do not seem to be entirely consistent with neither each other nor with perturbation theory.
\newline\indent
The Swiss cheese models used in \cite{syksyCMB} and \cite{Ishak} differ from each other in significant ways aside from the anisotropy issue. Hence, a comparison between the results in those two papers cannot yield much insight into the possible effects anisotropic structure formation has on observations. The purpose with the work presented here is to obtain such information, {\em i.e.} the study presented here is concerned with determining if the anisotropy of quasispherical Szekeres structures affects observations such as $D_{A,ls}$ and whether or not the anisotropy of the Szekeres models influences (dis-)agreement with perturbative results. The study is conducted by constructing four different Swiss cheese models that only differ in the shape of their individual structures. The structures in two of the models are anisotropic Szekeres structures, while their underlying LTB models are used for constructing the other two. By studying Swiss cheese models with both anisotropic structures and their ``underlying" spherically symmetric structures, it is possible by direct comparisons to quantify effects of anisotropy. Besides computing the average of $D_{A,ls}$ in each of the models, the density, the distance-redshift relation, the shear and the expansion rate along individual rays are also studied.

\section{Model specifications}\label{sec:models}
This section serves to introduce the Szekeres models and specify the particular models used in this study. In addition, a brief description of the construction of Swiss cheese models is given.
\newline\newline

\begin{figure*}
\centering
\subfigure[]{
\includegraphics[scale = 0.8]{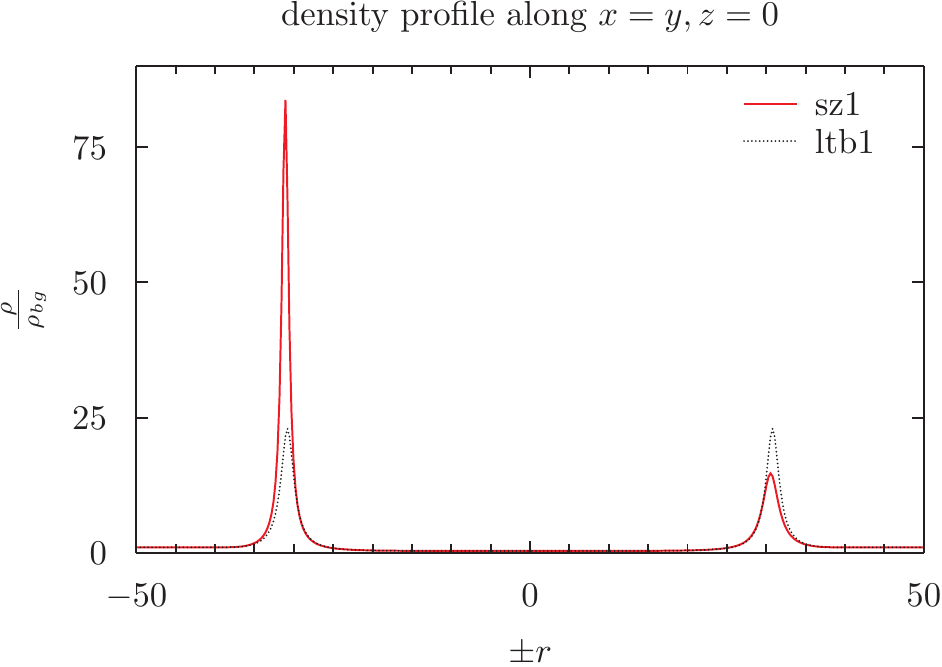}
}
\subfigure[]{
\includegraphics[scale = 0.8]{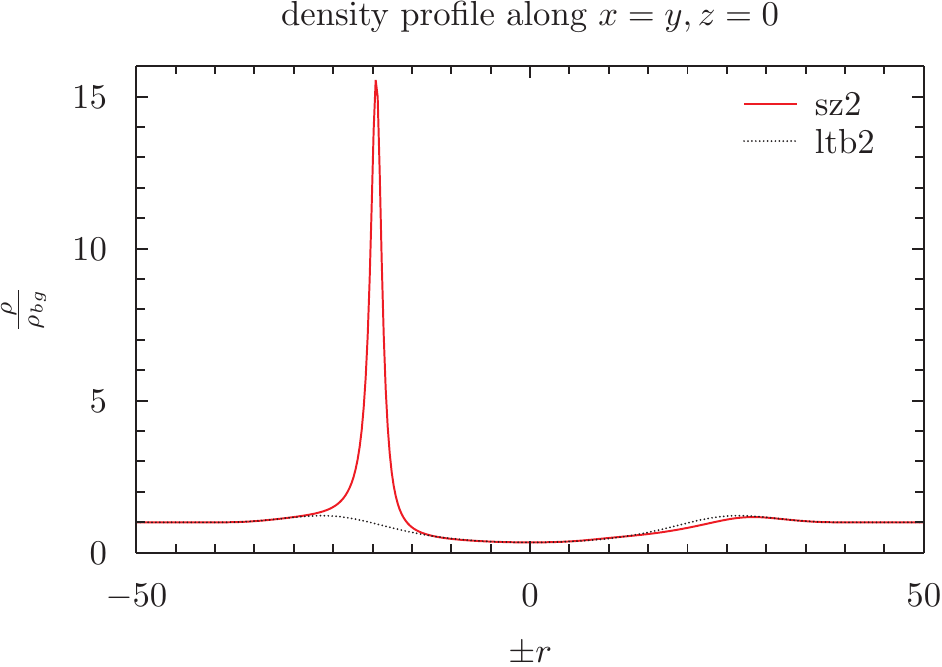}
}
\caption{Present time 1D density profiles of anisotropic Szekeres models (solid lines) and their underlying LTB models (dotted lines). Models sz1 and ltb1 are shown in the figure to the left while models sz2 and ltb2 are shown in the figure to the right. Comoving coordinates are in units of Mpc.}
\label{fig:densities}
\end{figure*}

The quasispherical Szekeres models \cite{Szekeres,Szekeres2} are exact, inhomogeneous dust solutions to Einstein's equations which can easily incorporate a homogeneous component with pressure \cite{Szafron} such as a cosmological constant. In general, the quasispherical Szekeres models have no Killing vectors \cite{killing} and hence no symmetries (see {\em e.g.} also \cite{killing_new}). Their line element can be written as:
\begin{equation}
\begin{split}
ds^{2} = -c^2dt^{2} +\frac{\left(A_{,r}(t,r)-A(t,r)\frac{E_{,r}(r,p,q)}{E(r,p,q)}\right)^2}{1-k(r)}dr^2\\ +
\frac{A(t,r)^2}{E(r,p,q)^2}(dp^2+dq^2)
\end{split}
\end{equation}
Subscripted commas followed by a coordinate or its index indicates partial derivative with respect to that coordinate.
\newline\indent
The metric function $E$ is given by $E = \frac{1}{2S}(p^2+q^2)-\frac{p P}{S} - \frac{qQ}{S}+\frac{P^2+Q^2+S^2}{2S}$, where $S, P$ and $Q$ are continuous but otherwise arbitrary functions of $r$ (with $S\neq 0$). The functions $P$, $Q$ and $S$ represent dipole distortions of the spacetimes of ``underlying" spherically symmetric LTB models. If they are chosen to be constant, the particular quasispherical Szekeres model reduces to such an underlying LTB model. In addition, the quasispherical Szekeres models have the FLRW models as their homogeneous and isotropic limit. The FLRW limit is achieved by setting $E = \frac{1}{2}\left(1+p^2+q^2 \right)$, $k(r)\propto r^2$ and $A(t,r) = ar$, with $a$ being the scale factor of the FLRW model. In the following, quasispherical Szekeres models will be referred to simply as Szekeres models. Since LTB models are also strictly speaking Szekeres models, Szekeres models which are specifically not spherically symmetric are sometimes referred to as anisotropic Szekeres models.
\newline\indent
For a Szekeres metric combined with dust and a cosmological constant, the diagonal components of Einstein's equation yield ($G_N$ denotes Newton's constant):
\begin{equation}\label{eq:A}
\frac{1}{c^2}A_{,t}^2 = \frac{2M}{A} - k +\frac{1}{3c^2}\Lambda A^2
\end{equation}
\begin{equation}
\rho = \frac{2M_{,r} - 6M\frac{E_{,r}}{E}}{c^2\beta A^2\left(A_{,r} - A\frac{E_{,r}}{E} \right) },\, \, \, \, \, \, \, \, \beta = 8\pi G_N/c^4
\end{equation}
The curvature function, $k$, and the integration constant, $M$, both depend on the radial coordinate only while $A$ depends on both $r$ and $t$. The top equation is also valid for LTB models, while the expression for the density in the LTB limit reduces to $\rho_{\text{LTB}} = \frac{2M_{,r}}{c^2\beta A^2A_{,r}}$ because $E_{,r} = 0$ in the limit of spherical symmetry.
\newline\newline
\begin{table}[]
\centering
\begin{tabular}{c c c c c c}
\hline\hline
Model & $n$ & $m$ & $Q_{\text{max}}$ & $n_p$ & $m_p$ \\
\hline
sz1/ltb1 &  6 & 6 & $1.768\cdot 10^{-5}$ & 10  & 4 \\
sz2/ltb2 &  2 & 4 & $4.368\cdot 10^{-3}$ & 2 & 10 \\
\hline
\end{tabular}
\caption{Specification of anisotropic Szekeres models and their underlying LTB models. $n$ and $m$ specify $k(r)$ and hence the LTB models while $Q_{\text{max}}, n_p$ and $m_p$ specify the functions $Q = P$ of the anisotropic Szekeres models.}
\label{table:models}
\end{table}
In general, neighboring voids and walls in the real universe are not expected to mass-compensate each other completely. Therefore, the most realistic void-wall Szekeres structures are presumably not those which reduce exactly to an FLRW model at a finite radius. However, by using such models, it is possible to obtain a larger packing fraction of voids in the Swiss cheese models without introducing discontinuities at boundaries between individual structures and the FLRW spacetime patches between the structures. Consequently, the models studied here are constructed so that they at $r = 40$Mpc reduce exactly to the flat $\Lambda$CDM model specified by $\Omega_{m,0} = 0.3$ and $H_0 = 70$km/s/Mpc. The comoving radius of $40$Mpc of the void-wall double-structures leads to a present day void radius slightly below $40$Mpc for two of the models and approximately $30$Mpc for the two other models.
\newline\indent
Reported typical sizes of voids in the Universe depend highly on the employed void definition including {\em e.g.} limits on void depths (compare {\em e.g.} the findings of \cite{voids_2001} and references therein with those of \cite{Voids_2002,voids_2003}). A void radius of approximately $40$Mpc fits well within the (broad) range of void sizes of \cite{voids_2001,voids_2003,Voids_2002} and with the voids used in other Swiss cheese studies including \cite{syksyCMB,Ishak}.
\newline\newline
The LTB models have two free functions and an extra degree of freedom from coordinate covariance in $r$. The coordinate covariance is here removed by setting $A(t_i, r) = a(t_i)r$, where $t_i$ is the age of the Universe at $z = 1200$ according to the $\Lambda$CDM background model specified above, and $a$ is the scale factor of that model. The first free function which is specified is the time of the big bang, {\em i.e.} the time where $A(t,r) = 0$. It is here set equal to zero so that the models contain no decaying modes (any constant value of the big bang time is equally valid for this purpose - see \cite{growing_modes} for details). At early times where the cosmological constant can be ignored, a big bang time of zero implies the following relation between $k$ and $M$ (see equation 2.14 in \cite{Acoleyen}):
\begin{equation}
M(r) = \frac{4\pi G_N\rho_{bg}(t_i)}{3c^2\left( a(t_i)r\right)^3 }\left(1 + \frac{3}{5}\frac{k(r)c^2}{\left(ra(t_i)H(t_i) \right)^2 } \right)
\end{equation}
$a, H$ and $\rho_{bg}$ denote the scale factor, Hubble parameter and density of the background model.
\newline\indent
The final specification of the LTB models is given by specifying $k$:
\begin{equation}\label{eq:k}
k(r) = \left\{ \begin{array}{rl}
-5.4\cdot 10^{-8}r^2\left(\left(\frac{r}{r_b} \right)^n -1 \right)^m  &\text{if} \,\, r<r_b \\
0 &\mbox{ otherwise}
\end{array} \right.
\end{equation}
The numerical values of the two constants $n$ and $m$ are indicated in table \ref{table:models}. $r_b$ is set equal to $40$Mpc.
\newline\newline
To further specify the Szekeres models, their dipole functions must also be given. Here, the choices $S = 1$ and $P = Q$ are made, with $P = Q$ specified as:
\begin{equation}\label{eq:Q}
P = Q = \left\{ \begin{array}{rl}
-r^2Q_{\text{max}}\left(\left(\frac{r}{r_b} \right)^{n_p} -1 \right)^{m_p}  &\text{if} \,\, r<r_b \\
0 &\mbox{ otherwise}
\end{array} \right.
\end{equation}
The numerical values of $Q_{\text{max}}, n_p$ and $m_p$ are shown in table \ref{table:models}.
\newline\newline
The four models are denoted sz1, ltb1, sz2 and ltb2. Present time density profiles of the models are shown in figure \ref{fig:densities}. The values of $n$ and $m$ have been chosen such that the resulting density contrasts are of similar size as those in the model studied in \cite{Ishak} and the void-profile used in \cite{syksyCMB}. However, the density contrasts of ltb1 and sz1 have been designed to be somewhat larger than the density contrasts of ltb2 and sz2 so that the effect of different density contrasts can be estimated. $Q_{\text{max}}, n_p$ and $m_p$ have been chosen so that large anisotropies are achieved while keeping in mind that a steeper transition between under- and overdensity leads to slower numerical computations. The models have also been designed so that the anisotropy of model sz1 is concentrated around the overdensity of the model while the anisotropy of model sz2 is clear also in its void-region.
\newline\indent
For simplicity, the Swiss cheese model based on model sz1 will also be referred to simply as sz1 and likewise for the other models.

\subsection{Construction of the Swiss cheese models}\label{sec:swisscheese}
The Swiss cheese models used here are constructed by arranging 1222 Szekeres structures with random (but fixed) orientations in a periodic box with side lengths of $860$Mpc. The Szekeres structures are arranged as a random close-packing of hard spheres of comoving radii approximately equal to $43$Mpc. The packing is obtained by using the Jodrey-Torey algorithm \cite{Jodrey,Jodrey2}. For appropriate initial conditions (see \cite{Jodrey}), the Jodrey-Torrey algorithm leads to a packing fraction of approximately $0.64$ which corresponds well with the experimentally obtained largest packing fractions of random close-packed hard spheres \cite{RCP_experimental}. Since the radius of the close-packed spheres used here is slightly larger than the radius of the actual Szekeres structures, the packing fraction of the Szekeres structures in the Swiss cheese models is only $0.515$. According to \cite{voids_77}, approximately $77\%$ of the Universe's volume is made up of voids. However, as mentioned in \cite{Voids_2002}, the ``packing fraction" of voids in the real universe depends on the employed definition of voids and survey analyses lead to observed packing fractions in the approximate range $0.3-0.75$ \cite{Voids_2002}, with, for instance, the packing fractions reported in \cite{Voids_2002,voids_2003} being roughly $0.3-0.4$. Hence, the packing fraction of approximately $0.5$ used here seems fairly reasonable. Note also that the packing fraction of $0.515$ lies in-between those used elsewhere: On one hand, the packing fraction used here is much larger than that used in other Swiss cheese models constructed as fixed spacetimes. For instance, in \cite{syksyCMB}, the actual packing fractions of voids is only approximately
\footnote{The packing fraction in \cite{syksyCMB} is given as approximately $0.34$ for spheres of comoving radius $50$Mpc/h. However, as mentioned in \cite{syksyCMB}, the LTB void model considered only deviates significantly from the FLRW backgrounds at $r \leq 30$Mpc/h. Hence, the actual packing fraction of LTB structures is only approximately $0.07$.}
$0.07$. On the other hand, most Swiss cheese models are not constructed as fixed spacetimes. Instead, LTB or Szekeres structures are arranged on-the-fly along light rays which makes it possible to obtain a quite large effective ``packing fraction" of voids along individual light rays. 
\newline\indent
A packing fraction larger than approximately $0.64$ in Swiss cheese models constructed as fixed spacetimes can be obtained by using multiple-sized structures. Random close-packing of di- and polydisperse spheres can {\em e.g.} be obtained with the algorithms of \cite{poly1,poly2}. Only Swiss cheese models with single-sized structures are considered here.
\newline\indent
A larger packing fraction can also be obtained by switching from a random close-packing to another close-packing distribution such as a cubic close-packing. However, the Swiss cheese models are more realistic when the distribution of structures are random. It should be noted though that even with a random close-packing og Szekeres structures, the distribution of matter in the Swiss cheese models is not particularly realistic and the models must be considered as merely toy-models of the Universe.
\newline\newline
All four Swiss cheese models have identical distributions, orientations and comoving sizes of structures so that the only difference between them is the shapes of their structures.

\section{Light propagation}\label{sec:light}
This section gives a very brief description of light propagation in Szekeres structures. For details, please see {\em e.g.} \cite{dig_selv2,Ishak2,Ishak}.
\newline\newline
In the geometric optics approximation, the paths of light rays can be obtained by solving the geodesic equations \cite{god_bog,lensing_bog},  $\frac{d}{d\lambda}\left(g_{\alpha\beta}k^{\beta} \right) = \frac{1}{2}g_{\beta\gamma,\alpha}k^{\beta}k^{\gamma} $, where $\lambda$ is an affine parameter, $g_{\alpha\beta}$ the metric tensor and $k^{\alpha}$ the null-geodesic tangent vector. In order to obtain the angular diameter distance along these light paths, it is necessary to describe the deviation between neighboring light rays in a light ray bundle. Following extensions of early considerations described in \cite{sachs}, this is done by using the transport equation \cite{arbitrary_spacetime}:
 \begin{equation}\label{D_dot}
\ddot D^a_b = T^a_cD^c_b
\end{equation}
$T_{ab}$ is the optical tidal matrix and has the following components:
\begin{equation}
 T_{ab} = 
  \begin{pmatrix} \mathbf{R}- Re(\mathbf{F}) & Im(\mathbf{F}) \\ Im(\mathbf{F}) & \mathbf{R}+ Re(\mathbf{F})  \end{pmatrix} 
\end{equation}
$\mathbf{R}: = -\frac{1}{2}R_{\mu\nu}k^{\mu}k^{\nu}$ and $\mathbf{F}:=-\frac{1}{2}R_{\alpha\beta\mu\nu}(\epsilon^*)^{\alpha}k^{\beta}(\epsilon^*)^{\mu}k^{\nu}$, where $R_{\mu\nu}$ is the Ricci tensor, $R_{\alpha\beta\mu\nu}$ the Riemann tensor and $\epsilon^{\mu} := E_1^{\mu} + iE_2^{\mu}$ with $E_1^{\mu}, E_2^{\mu}$ spanning screen space.
\newline\indent
$D^a_b$ in equation (\ref{D_dot}) is the deformation tensor which describes the evolution of the shape of the screen space area of the given light ray bundle (see {\em e.g.} \cite{arbitrary_spacetime} for details). In particular, $\sqrt{|\det D^a_b|}$ describes the angular diameter distance along the light ray when initial conditions are set appropriately (see \cite{dig_selv2} for initial conditions in the case of Szekeres models). The angular diameter distance along an individual light ray is therefore obtained by solving the transport equation simultaneously with the geodesic equations and the equations of parallel transport of $E_1^{\mu}, E_2^{\mu}$ along the given null-geodesic.
\newline\newline
The results regarding the angular diameter distance will in the following section be represented by the shift, $\delta D_A :=\frac{D_A-D_{A,bg}}{D_{A,bg}}$, in the angular diameter distance, where $D_A$ is the angular diameter distance computed along a given light ray while $D_{A,bg}$ is the corresponding angular diameter distance in the background model at the same redshift. In standard first-order perturbation theory, fluctuations in $D_A$ are described by the convergence $\kappa$ according to the relation $D_A \approx D_{A,bg}\left(1 -\kappa \right) $, {\em i.e.} $\delta D_A\approx - \kappa$. As mentioned in \cite{kappa_explained}, in first-order perturbation theory, the convergence can be split into five components, namely the ISW and SW contributions, the Shapiro time-delay contribution, the Doppler contribution and the gravitational convergence (see {\em e.g.} \cite{bonvin} for a derivation). The first three contributions are subdominant to the other two. The Doppler convergence arises due to redshift perturbations. In standard perturbation theory based on the Newtonian gauge, the Doppler convergence can be ascribed peculiar motions of the source and observer. In LTB and Szekeres models described in a comoving foliation of spacetime there are no peculiar motions but the Doppler convergence persists since the inhomogeneous spacetime induces fluctuations in the redshift. The Doppler convergence is a local effect that becomes subdominant to the gravitational convergence at redshifts $z\gtrsim 0.5$ (see {\em e.g.} \cite{cosmo_doppler,dig_selv2}). The gravitational convergence is given by  $\kappa_{\delta} = \frac{4\pi G_N}{c^2}\int_{0}^{r_s}dra^2\delta\rho\frac{(r_s-r)r}{r_s}$, where $\delta\rho :=\rho-\rho_{bg}$, $r_s$ is the radial distance from the observer (placed at the origin) to the source, and the integral is along the Born approximated light path. Clearly, the primary contributions to $\kappa_{\delta}$ come from low redshift structures as density fluctuations at early times are suppressed compared to those at late times.
\newline\indent
Since $\kappa \approx -\delta D_A$, these considerations also apply to $\delta D_A$. It should therefore not be necessary to populate the entire universe along a light ray with structures in order to obtain the main parts of the effects inhomogeneities have on light propagation. Although all light rays are traced back to the time of last scattering, here defined
\footnote{The surface of last scattering could also be defined as that corresponding to $t = t_{ls}$. The difference between the two definitions is negligible and at any rate sub-dominant to effects arising in the real universe due to decoupling not actually being instantaneous.}
by $z = 1100$, light rays are therefore not entered into any new structures once the redshift exceeds seven, {\em i.e.} for $z\geq7$. By studying a few individual light rays it has been checked that this is sufficient to obtain the main part of the effects that inhomogeneities have on light propagation. Note though, that primary anisotropies of the CMB are due to inhomogeneities at the last scattering surface which are not included here.
\newline\newline
As implied in the above discussion of the Doppler convergence, the redshift along light rays in the Swiss cheese models will fluctuate around the background redshift. As discussed in \cite{syksy_av1,syksy_av2}, these fluctuations should not accumulate over large distances since the Swiss cheese models studied here are statistically homogeneous and isotropic with structures only evolving a small amount during the time it takes a light ray to traverse them. This assertion is in agreement with studies based on redshift fluctuations in Swiss cheese models such as \cite{LTB_in_NG3} which in fact show that the accumulation in the redshift fluctuation from even a single structure is typically negligible. The results of section \ref{subsec:expansion_shear} show that this is also the case for the models studied here.

\section{Results}\label{sec:results}
In this section, the results obtained from studying light propagation in the four Swiss cheese models are presented. Just as the four models are identical except for the shapes of the individual structures, the position of the observer (placed randomly in an FLRW patch) and the lines of sight are the same in the four studies.
\newline\indent
As part of the scheme used to distribute Szekeres structures in the Swiss cheese models, each structure was given a number. If two light rays in different models travel through the same sequence of structures, they will be said to travel through equivalent portions of spacetime.

\subsection{Individual light rays}
\begin{figure*}
\centering
\subfigure[]{
\includegraphics[scale = 0.7]{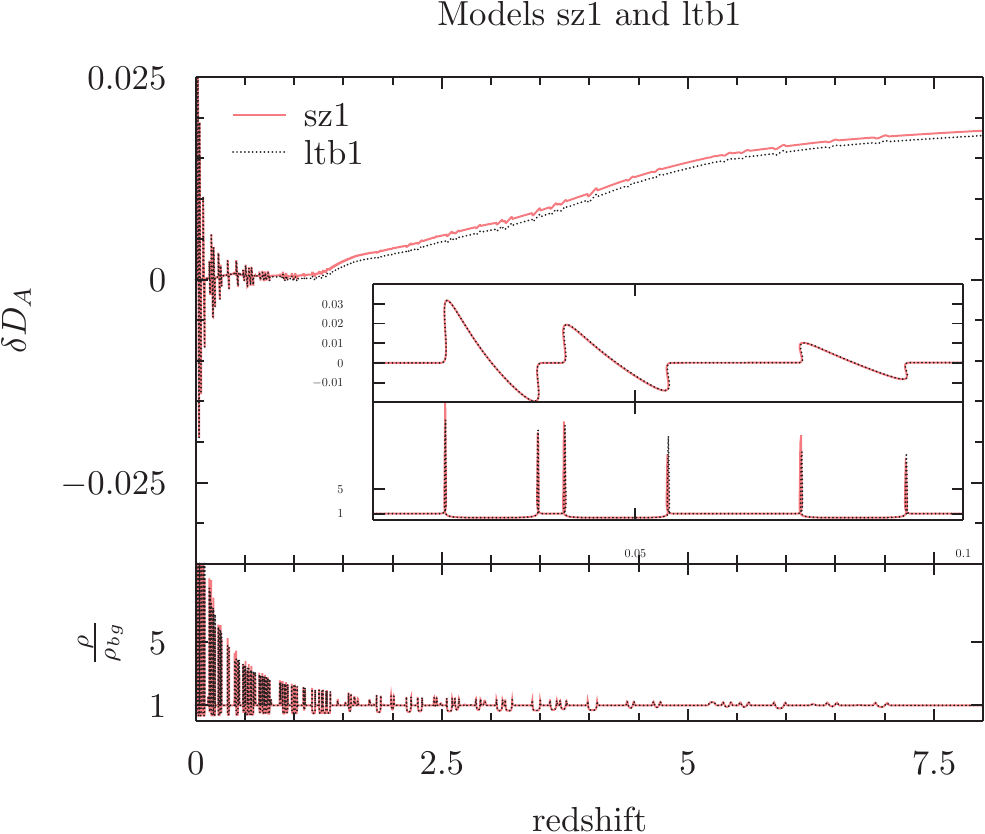}\label{subfig:dDA_1}
}
\subfigure[]{
\includegraphics[scale = 0.7]{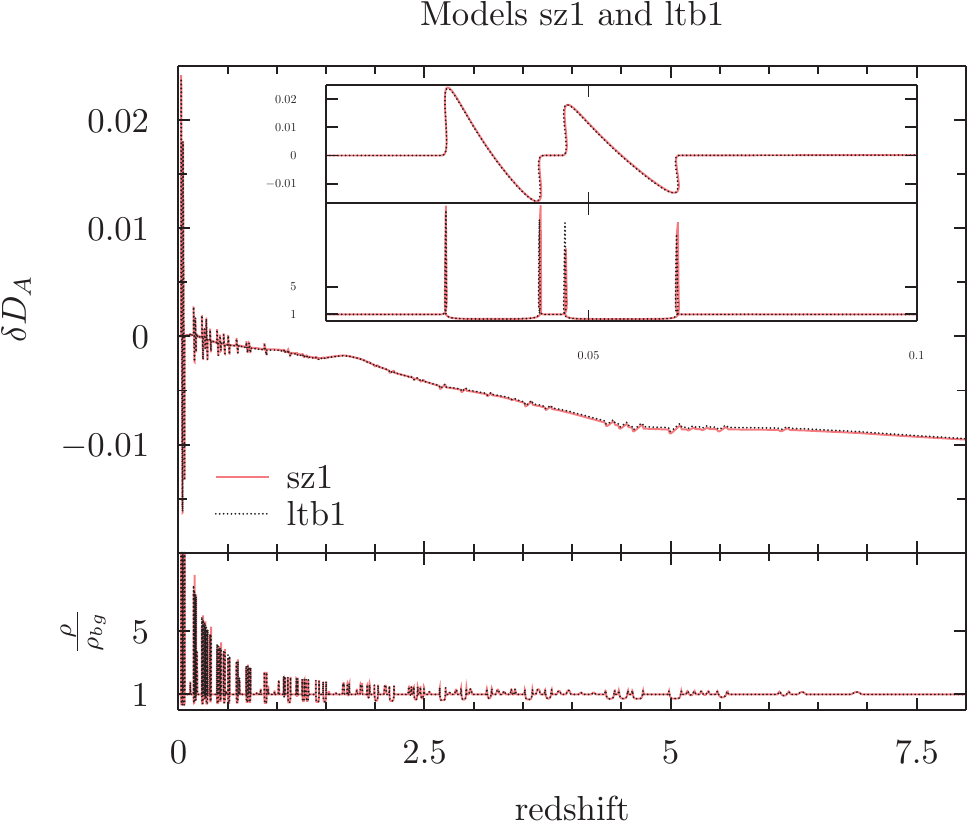}\label{subfig:dDA_1_neg}
}\par
\subfigure[]{
\includegraphics[scale = 0.7]{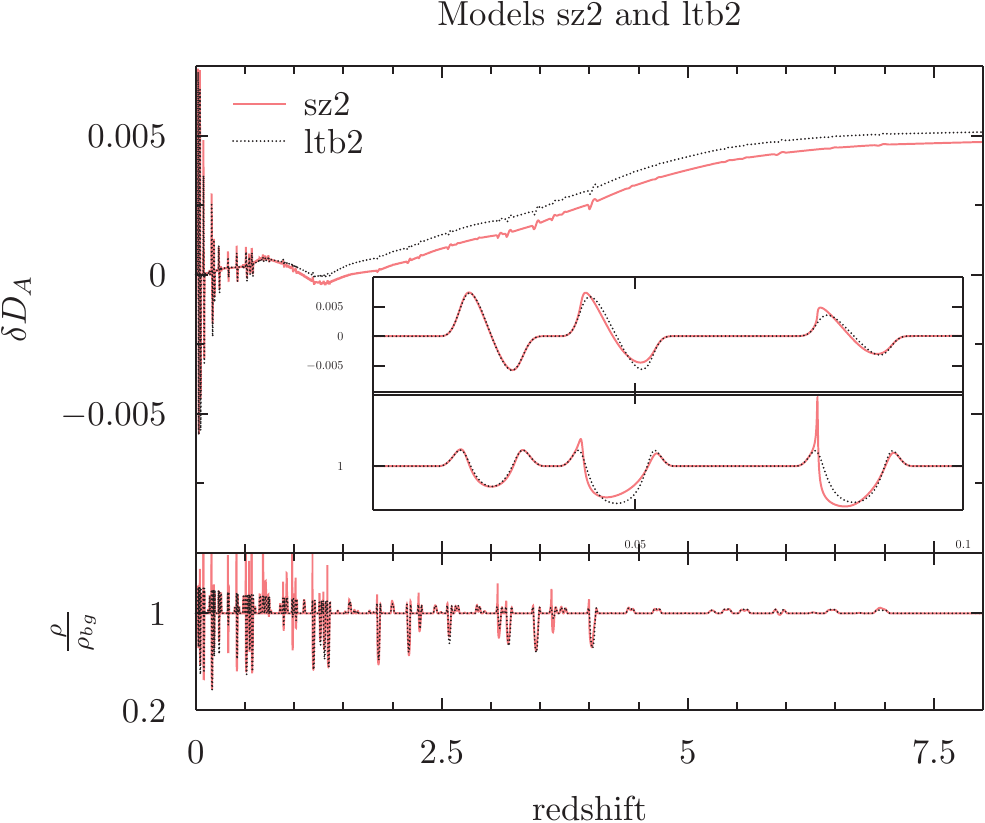}\label{subfig:dDA_2}
}
\subfigure[]{
\includegraphics[scale = 0.7]{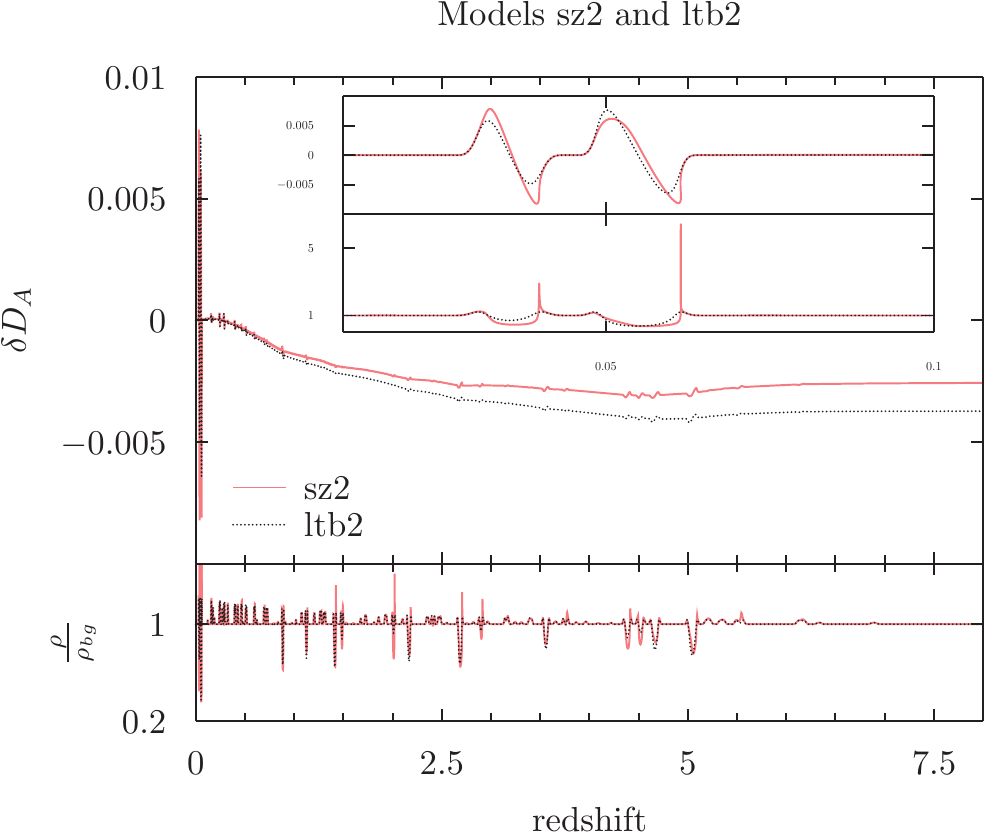}\label{subfig:dDA_2_neg}
}
\caption{The shift in the angular diameter distance along individual light rays and the corresponding densities. Results along equivalent rays in models based on Szekeres and underlying LTB structures are shown in the same figure with solid and dotted lines, respectively. The results are shown for two light rays in each model, namely one with a negative and one with a positive $\delta D_A$ at high redshifts. The light rays corresponding to subfigures a and c are equivalent while the same is true for those corresponding to subfigures b and d. The plots are only shown in the interval $z\in[0,8]$ since the light rays are set to not enter any new structures for $z\geq7$. Close-ups are included for the redshift interval $z\in\left[0.01:0.1 \right]$.}
\label{fig:dDA}
\end{figure*}

For each model, the redshift-distance relation along two individual light rays has been computed. The results are shown in figure \ref{fig:dDA}. Note that the figures clearly show that the gravitational convergence becomes non-negligible around $z = 0.1$ and becomes dominant before $z = 1$. As discussed in the previous section, this is exactly as expected and has earlier been studied in \cite{dig_selv2,cosmo_doppler} (see also \cite{bright_side,bonvin}).
\newline\indent

Figure \ref{fig:dDA} shows the results along two groups of four light rays initialized identically but each in a different Swiss cheese model. Within these two groups, all four light rays turn out to travel through equivalent portions of spacetime. While it is not too surprising that the light rays of models ltb1 and ltb2 have equivalent paths, the result is less obvious for models sz1 and sz2 since light rays are bent significantly by the anisotropic structures of these models. Apparently, the bending of a light ray when it enters a structure is largely countered by the bending it experiences on its way out of the structure. It is not expected that such a result is valid in a general spacetime, {\em i.e.} if two general inhomogeneous spacetimes have the same distribution of (differently shaped) structures, there is no reason to expect that the null geodesics of the two spacetimes are equivalent. Instead, the result found here is expected to be due to the specific dipole nature of the anisotropy of the Szekeres spacetimes. Indeed, the anisotropy of Szekeres models is very special as it renders the spacetime outside the inhomogeneity entirely unaffected by the structure despite the general Szekeres structures not being spherically symmetric. This is an important feature of the Szekeres models as it is necessary in order for the models to reduce to FLRW models outside the structures and hence be appropriate for Swiss cheese constructions. Unfortunately, it seems that this same feature limits the extent to which effects of anisotropy on light propagation can be studied with these models.
\newline\newline
As illustrated in figure \ref{fig:dDA}, although the light rays travel through equivalent portions of spacetime, their redshift-distance relations are not identical. This is because the density distributions along the light rays are not identical and hence {\em e.g.} the gravitational convergence will not be the same along the individual rays. The smaller density fluctuation amplitudes of models sz2 and ltb2 compared to those of models sz1 and ltb1 therefore explain the smaller numerical values of $\delta D_A$ along light rays in the former two models.
\newline\indent
While $\delta D_A$ is roughly the same in models sz1 and ltb1, there is a clear difference between $\delta D_A$ along the rays in models sz2 and ltb2. This could be a result of the fact that the anisotropy of model sz1 is most prominent at the overdensities of the structures while the anisotropy of the structures in model sz2 is significant also inside the voids. Indeed, the densities along the light rays in models sz1 and ltb1 are almost identical with only a small difference at the overdensities. Contrary to this, there is a clear difference between the densities along the light rays in models sz2 and ltb2. This can be seen in the close-ups of the density fields shown in figure \ref{fig:dDA}.

\subsubsection{Expansion rate and shear}\label{subsec:expansion_shear}
\begin{figure*}
\centering
\subfigure[]{
\includegraphics[scale = 0.8]{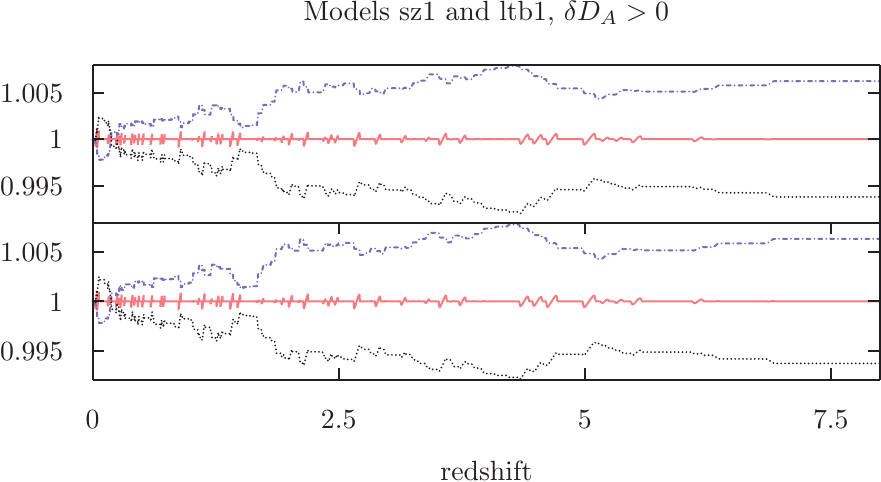}\label{subfig:shear1_sz1}
}
\subfigure[]{
\includegraphics[scale = 0.8]{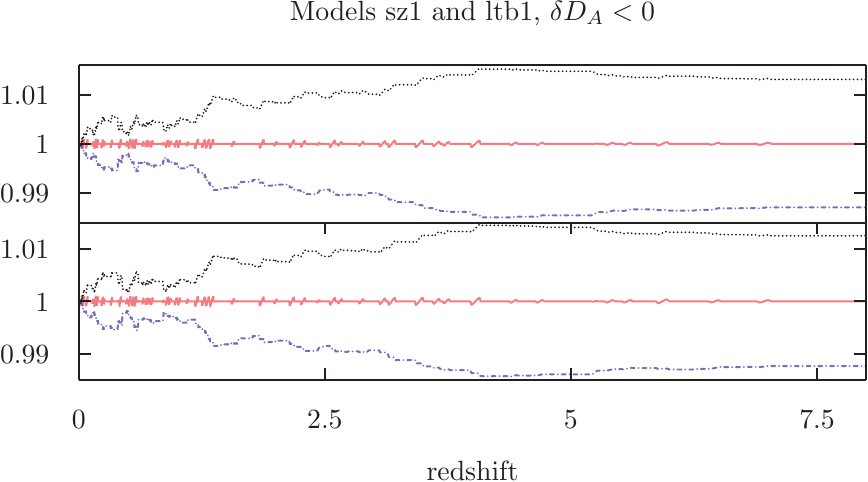}\label{subfig:shear1_sz1_neg}
}\par
\subfigure[]{
\includegraphics[scale = 0.8]{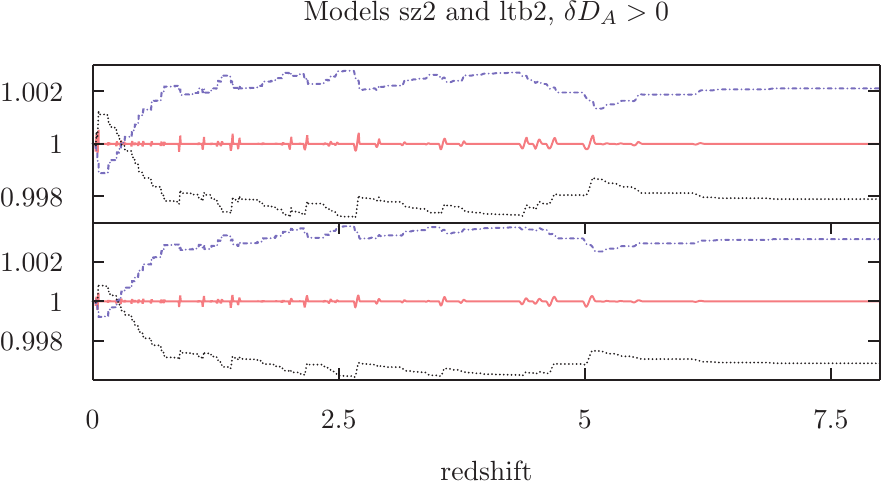}\label{subfig:shear1_sz2}
}
\subfigure[]{
\includegraphics[scale = 0.8]{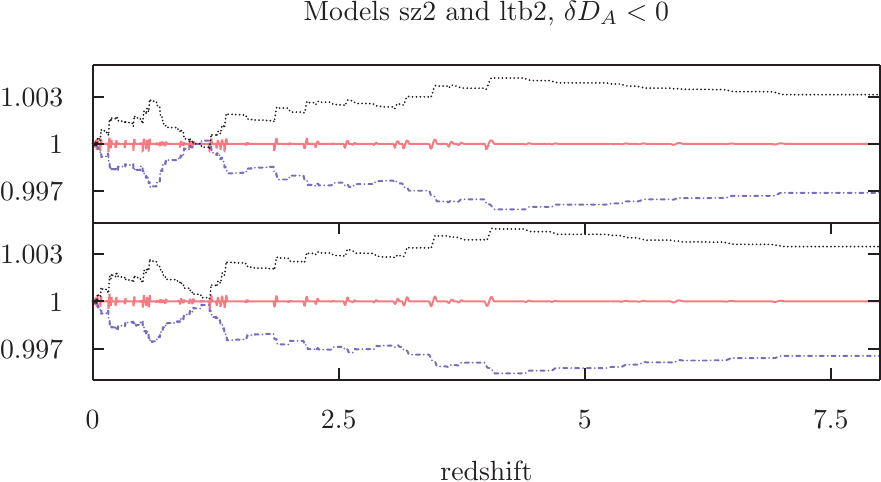}\label{subfig:shear1_sz2_neg}
}\par
\includegraphics[scale = 0.8]{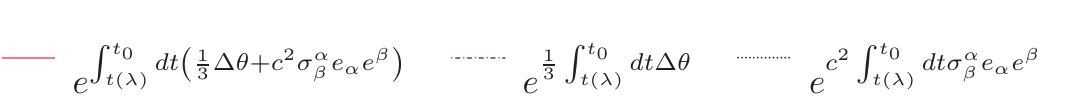}\label{subfig:captions}
\caption{The exponential of the integrals of $\Delta\theta$ and $\sigma^{\alpha}_{\beta}e_{\alpha}e^{\beta}$. In each figure, the upper plots are for Szekeres models while the lower plots are for the corresponding LTB models. Solid lines show $e^{\int_{t(\lambda)}^{t_0}dt \left(\frac{1}{3}\Delta\theta + c^2\sigma^{\alpha}_{\beta}e_{\alpha}e^{\beta} \right) }$ while dash-dotted lines show $e^{\frac{1}{3}\int_{t(\lambda)}^{t_0}dt \Delta\theta }$ and dotted lines show $e^{c^2\int_{t(\lambda)}^{t_0}dt \sigma^{\alpha}_{\beta}e_{\alpha}e^{\beta} }$. Figure titles indicate whether the given light ray has a positive or negative shift in $D_A$ at high redshifts.}
\label{fig:shear1}
\end{figure*}
The redshift along a light ray can be computed as \cite{syksy_av1,syksy_av2} (see {\em e.g.} also \cite{god_bog,Ehlers}):
\begin{equation}\label{eq:z}
\begin{split}
1+z = e^{\int_{t(\lambda)}^{t_0}dt\left(\frac{1}{3}\theta + c^2\sigma_{\alpha}^{\beta}e^{\alpha}e_{\beta} \right)  }\\ = e^{\int_{t(\lambda)}^{t_0}dt\left(\frac{1}{3}\left\langle \theta\right\rangle  \right)  }\cdot e^{\int_{t(\lambda)}^{t_0}dt\left(\frac{1}{3}\Delta\theta + c^2\sigma_{\alpha}^{\beta}e^{\alpha}e_{\beta} \right)  }
\end{split}
\end{equation}
$\theta:=u^{\alpha}_{;\alpha} = \left\langle\theta \right\rangle + \Delta\theta = 3H + \Delta\theta $ denotes the local expansion rate while  $\sigma_{\alpha\beta} = u_{\left(\alpha;\beta \right) } - \frac{1}{3}h_{\alpha\beta}\theta$ is the shear with $h_{\alpha\beta} = g_{\alpha\beta} + u_{\alpha}u_{\beta}$ the projection tensor projecting onto hypersurfaces orthogonal to the dust velocity field $u^{\alpha}$. $e^{\alpha} = \frac{\frac{u^{\alpha}}{c^2}-k^{\alpha}}{u^{\beta}k_{\beta}}$ is proportional to the spatial direction of $k^{\alpha}$. For the Szekeres metric, $\theta$ and the non-vanishing components of $\sigma^{\alpha}_{\beta}$ are:
\begin{equation}
\begin{split}
\theta = \frac{ A_{,tr} -3A_{,t} \frac{E_{,r}}{E} + 2\frac{A_{,t}A_{,r}}{A} }{ A_{,r} -A\frac{E_{,r}}{E}  }\\
\sigma^{r}_{r} = \frac{2}{3}\frac{A_{,tr} - \frac{A_{,t}A_{,r}}{A}}{A_{,r} - A\frac{E_{,r}}{E}}\\
\sigma^p_p = \sigma^q_q = -\frac{1}{2}\sigma^r_r
\end{split}
\end{equation}
It was argued in \cite{syksy_av1,syksy_av2} that the contributions of $\Delta \theta$ and $\sigma_{\alpha\beta}e^{\alpha}e^{\beta} $ should vanish up to statistical fluctuations in the integral of equation (\ref{eq:z}) for statistically homogeneous and isotropic spacetimes with slowly evolving structures. However, for the Swiss cheese model studied in \cite{tardis}, it was found that the integrals of $\Delta\theta$ and the projected shear did not cancel individually, but they did cancel with each other. As noted in \cite{tardis}, it is interesting to see if this feature also appears for models without surface layers. The two quantities have therefore also been considered along the eight studied individual light rays. The results are shown in figure \ref{fig:shear1}. The figure clearly shows that the contributions from $\Delta \theta$ and $\sigma_{\alpha\beta}e^{\alpha}e^{\beta} $ cancel with each other to a high precision. It would be very interesting to learn if this is a feature particular to the Szekeres models or if it is a more general result.
\newline\indent
As with the angular diameter distance results, the results regarding $\Delta \theta$ and $\sigma_{\alpha\beta}e^{\alpha}e^{\beta} $ are very similar for models sz1 and ltb1, and sz2 and ltb2, with the difference between the results of the two latter being slightly more prominent.

\subsection{The line-of-sight averaged angular diameter distance to the last scattering surface}

\begin{table*}[]
\centering
\begin{tabular}{c c c c c}
\hline\hline
Model & $\left\langle \delta D_{A,ls}\right\rangle $ & $\sigma$ & $95 \%$ limits for $\left\langle \delta D_{A,ls}\right\rangle $& $99 \%$ limits for $\left\langle \delta D_{A,ls}\right\rangle $  \\
\hline
sz1 & $-1.13\cdot 10^{-4}$  &  $5.43\cdot 10^{-5}$ & $\left[-2.19\cdot 10^{-4} ,-6.81\cdot 10^{-6}\right]$  & $\left[-2.52\cdot 10^{-4},2.73\cdot 10^{-5} \right] $ \\
ltb1 & $-1.10\cdot 10^{-4}$ & $5.42\cdot 10^{-5}$ & $\left[ -2.16\cdot 10^{-4}, -4.09\cdot10^{-6} \right] $ & $\left[ -2.48\cdot 10^{-4},3.00\cdot 10^{-5} \right] $\\
sz2 & $-1.60\cdot 10^{-6}$ &$1.81\cdot 10^{-5}$ & $\left[ -3.73\cdot 10^{-5},3.38\cdot 10^{-5}  \right] $  &  $\left[ -4.81\cdot 10^{-5},4.51\cdot 10^{-5}  \right] $\\
ltb2 & $1.78\cdot 10^{-6}$ & $1.79\cdot 10^{-5}$ & $\left[ -3.32\cdot 10^{-5}, 3.69\cdot 10^{-5}\right] $ & $ \left[ -4.41\cdot 10^{-5},4.78\cdot 10^{-5} \right]  $\\
\hline
\end{tabular}
\caption{Average shift in the angular diameter distance to the last scattering surface, $\left\langle \delta D_{A,ls}\right\rangle $, together with a standard deviation and $95\%$ and $99\%$ confidence interval obtained by bootstrapping with $10^{5}$ samples.}
\label{table:dDA}
\end{table*}

\begin{figure*}
\centering
\subfigure[]{
\includegraphics[scale = 0.9]{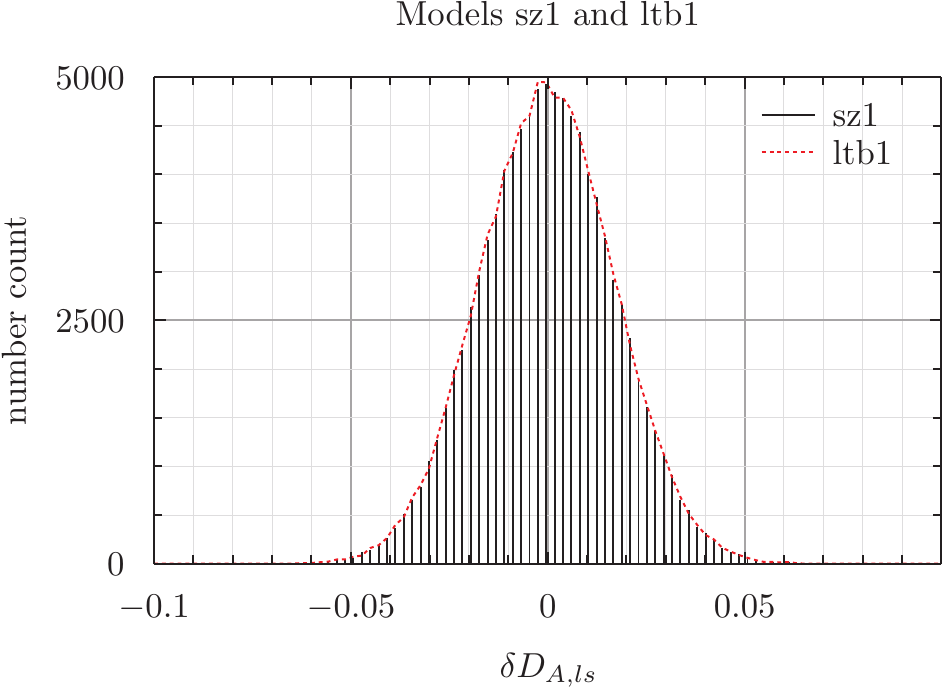}
}
\subfigure[]{
\includegraphics[scale = 0.9]{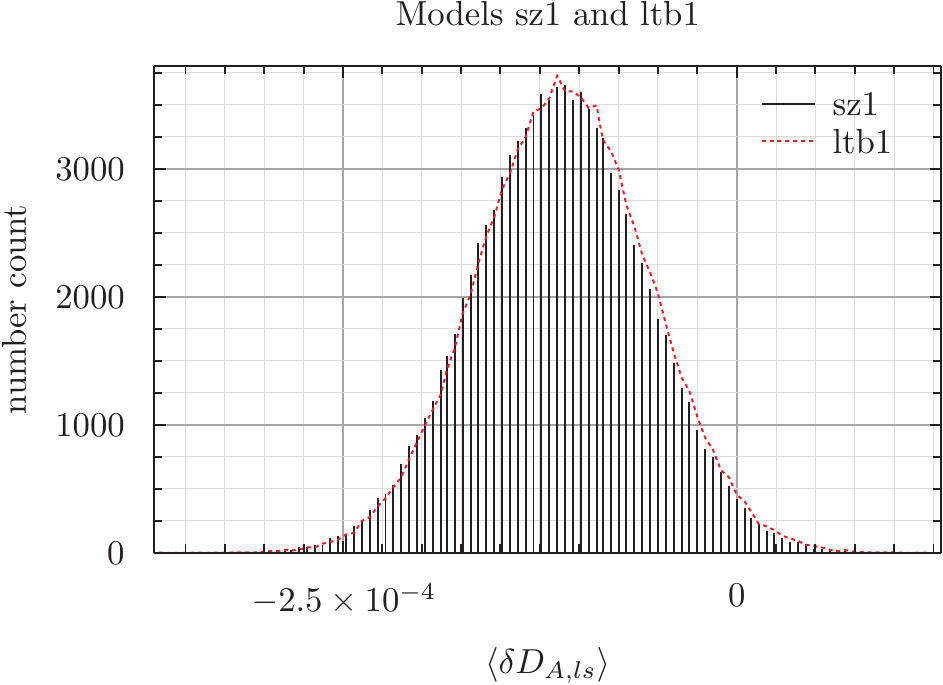}
}\par
\subfigure[]{
\includegraphics[scale = 0.9]{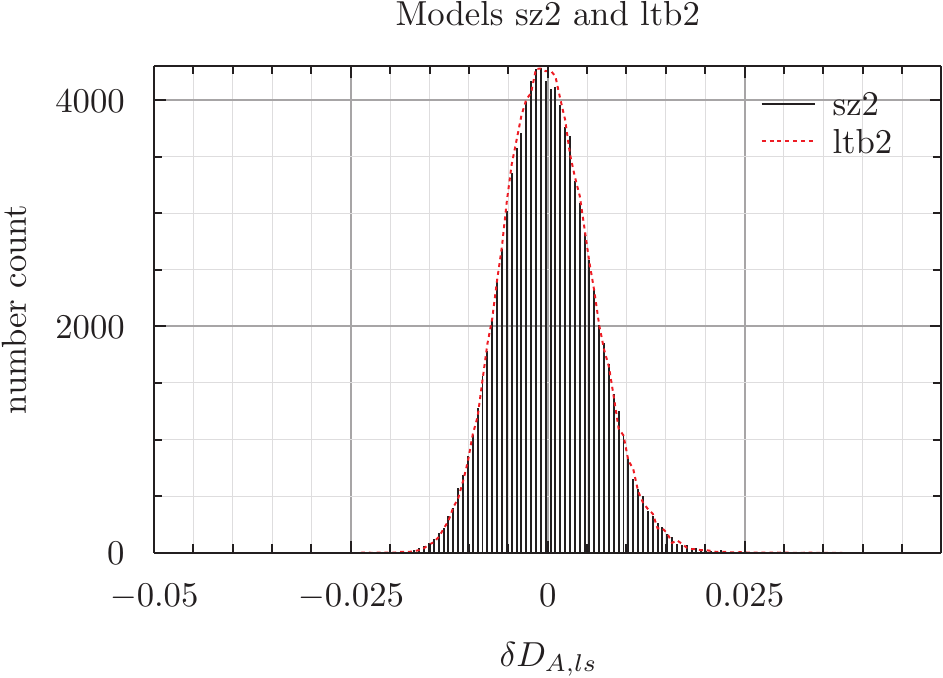}
}
\subfigure[]{
\includegraphics[scale = 0.9]{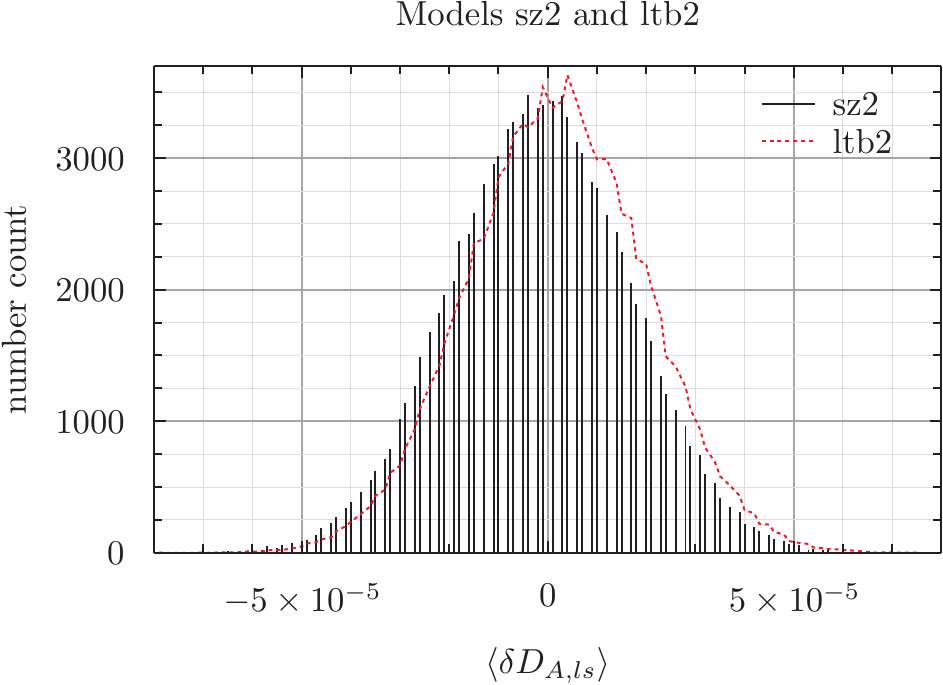}
}
\caption{Distributions of $\delta D_A$ and bootstrap distributions of $\left\langle \delta D_A \right\rangle $ obtained for the four models. Results for sz1 and sz2 are shown as bar charts with the bars centered at each subinterval. The results for ltb1 and ltb2 are shown with dotted lines.}
\label{fig:bootstrap}
\end{figure*}

In figures \ref{subfig:dDA_1} and \ref{subfig:dDA_2}, $\delta D_A$ is positive at high redshifts while it is negative in figures \ref{subfig:dDA_1_neg} and \ref{subfig:dDA_2_neg}. The feature decisive for the sign of $\delta D_A$ is the ``amount" of over- and underdensity along the light rays, with the negative $\delta D_A$ achieved when overdensities dominate sufficiently along a given light ray
\footnote{This is true regardless of the contribution from the Weyl tensor as this always leads to a decrease in the angular diameter distance compared to the background. This is {\em e.g.} seen by equation 2.8a in \cite{arbitrary_spacetime} and is also discussed in {\em e.g.} \cite{lensing_bog,Kaiser_Peacock}.}
. For the real universe, it is expected that light rays spend most of their time in underdense regions and hence a positive $\delta D_A$ is more common. In the Swiss cheese models studied here however, the light rays cannot enter into voids without also moving through compensating overdensities. On the other hand, it {\em is} possible for the light rays to move through overdense parts of the individual structures without entering into the underdense regions (this was also pointed out in \cite{LTB_in_NG3}). Therefore, it is {\em a priori} expected that a negative $\delta D_A$ should be more common in the Swiss cheese models. Table \ref{table:dDA} shows the average shift in $D_{A,ls}$, $\left\langle \delta D_{A,ls}\right\rangle $, obtained for the four models studied here. $98304$ light rays have been used to compute the average for each model. Following the approach in \cite{syksyCMB}, the errors and confidence intervals have been obtained with a bootstrap approach (first introduced in \cite{bootstrap}), here based on $10^5$ samples. Distributions of $\delta D_{A,ls}$ and $\left\langle \delta D_{A,ls}\right\rangle $ are shown in figure \ref{fig:bootstrap}.
\newline\indent
For three of the four models studied here, the average shifts obtained are negative, corresponding to an average image magnification. As explained above, this makes sense intuitively. More importantly, it is in agreement with the analyses of {\em e.g.} \cite{angle_vs_ensemble} which shows that the angular diameter distance decreases upon line-of-sight averaging which is the type of averaging conducted here since fixed spacetimes are used. The corresponding $95\%$ confidence intervals obtained for models ltb1 and sz1 are purely negative while their $99\%$ confidence intervals contain zero and positive values. For model ltb2, already the $95\%$ confidence interval contains positive values. In addition, the average angular diameter distance of model ltb2 is itself positive. However, the shift is very small and in particular much smaller than the estimated error (corresponding to one standard deviation), as is also the case for model sz2. Even for models sz1 and ltb1, the negative shifts are only significant to $2\sigma$. It is notable that the results obtained here have such a low significance even though eight times more light rays were used here than in \cite{syksyCMB}, where a similar significance ($1\sigma$) was obtained for the void-model of that study. Also in \cite{Ishak} a much higher statistical significance was obtained with a much lower number of light rays. For models sz1 and sz2 this could be due to the larger density contrast of the models compared to those in \cite{syksyCMB,Ishak}. Models sz2 and ltb2 have density fluctuations closer to those in \cite{syksyCMB,Ishak}, but for these two models the lack of statistical significance is clearly a consequence of the small numerical value of the average.
\newline\newline
The dominant sign (negative) of the shifts obtained here is opposite of that found in most other Swiss cheese studies based on LTB and Szekeres models and in particular of those found in \cite{syksyCMB,Ishak} (see {\em e.g.} \cite{Nbody_LTB_average} for one of the few exceptions). However, as mentioned in the introduction, the average in \cite{Ishak} must be considered an ensemble average and not an line-of-sight average
\footnote{In the erratum of \cite{Ishak}, the authors classify their average as a line-of-sight average. The authors of \cite{Ishak} agree that it is in fact an ensemble average that they compute. For the purposes in \cite{Ishak}, the ensemble average was considered sufficient as an approximation of a line-of-sight average.}
; the results studied in \cite{Ishak} are obtained by tracing light rays along distributions of Szekeres models constructed on-the-fly and not in a fixed spacetime. Therefore, the averages of \cite{Ishak} are over different realizations of a light ray in a single direction, {\em i.e.} an ensemble average as described in {\em e.g.} \cite{angle_vs_ensemble,Kaiser_Peacock}. It was shown in \cite{angle_vs_ensemble} that the ensemble average of the angular diameter distance is increased compared to the background value (at second order). Therefore, the negative shift obtained here and the positive shift obtained in \cite{Ishak} are in agreement with each other and with second-order perturbation theory. The positive shift obtained in \cite{syksyCMB} is based on an line-of-sight average and ``should" therefore have been negative. However, the seeming discrepancy can possibly be attributed to the fact that the results of \cite{syksyCMB} are (like those presented here) statistically {\em in}significant. Indeed, due to technical details, the results obtained here were obtained using several sub-samples of light rays, including ones with $4/5$ of the total number of light rays studied per model in \cite{syksyCMB}. Several of these sub-samples had positive shifts while others had negative.
\newline\newline
As mentioned in the introduction, well-known estimates indicate that the ensemble of $\mu$ should equal $1$ \cite{weinberg} and that the line-of-sight average of $\mu^{-1}$ should also be equal to 1 \cite{angle_vs_ensemble_early}, both being true only if the area of the relevant constant-redshift surface is the same as in the background FLRW model (see also \cite{angle_vs_ensemble,Kaiser_Peacock}). The line-of-sight average value of $\mu-1$ and $\mu^{-1}-1$ together with confidence intervals and standard deviations for the models studied here are shown in table \ref{table:mu}. As seen, the results show good agreement with the expectation that $\left\langle \mu^{-1}\right\rangle -1\approx 0$. On the other hand, $\left\langle \mu\right\rangle -1$ is not likely to contain the value zero. This is also expected as it is instead the ensemble average of $\mu$ which should equal $1$ and not the line-of-sight average computed here. Thus, while the results obtained here are again consistent with perturbation theory, they are not consistent with the results of \cite{syksyCMB,Ishak}. In fact, the results obtained for the void-model of \cite{syksyCMB} are the opposite of those obtained here with the line-of-sight average of $\mu$ corresponding well with $1$ but the line-of-sight average of $\mu^{-1}$ being very unlikely to be consistent with the value $1$. The results discussed in the erratum of \cite{Ishak} indicate that the obtained ensemble average of $\mu^{-1}$ is consistent with $1$, again contrary to what is expected based on perturbation theory (if the total areas of constant redshift surfaces are largely the same as in the background FLRW models).
\newline\newline
The average values, confidence intervals and estimated errors discussed above are very similar for models sz1 and ltb1, and sz2 and ltb2. This is in good agreement with the results of the previous subsection and shows that the effects of the anisotropy of the Szekeres models have a quite small overall effect on light propagation results, at least when light rays traverse entire structures.
\newline\indent
Also as in the earlier subsections, the agreement between the results of models sz1 and ltb1 are more striking than the agreement between those of models sz2 and ltb2. In particular, the sign of $\left\langle \delta D_A\right\rangle $ are opposite indicating a poor agreement. However, the numerical values of $\left\langle \delta D_A\right\rangle $ for these two models are quite small and have a very low statistical significance. Note also that the positive value $\left\langle \delta D_A\right\rangle $ obtained for ltb2 can be understood in terms of the intuitive considerations at the beginning of this subsection, {\em i.e.} by noting that the overdensities surrounding the voids in ltb2 have quite small amplitudes. This means that effects on a light ray moving through overdensities without entering void regions are diminished compared to in the other three models. This feature also indicates that a particularly large sample size is needed in order to obtain a negative average shift (assuming that a negative shift is indeed the result obtained with the sample size going to infinity); as discussed in \cite{angle_vs_ensemble}, the fact that ensemble averages of $D_{A,ls}$ should be positive while its line-of-sight average should be negative indicates that it is more likely for a light ray to travel through underdensities while a few special light rays that travel more through overdensities have larger numerical values of their shifts so that they make up for (and more), for the positive shifts of the other light rays. 
\begin{table*}[]
\centering
\begin{tabular}{c c c c c}
\hline\hline
Model &  $99 \%$ limits for $\left\langle \mu^{-1}\right\rangle-1 $ &$\left\langle \mu^{-1}\right\rangle -1 \pm\sigma_{\mu^{-1}}$& $99 \%$ limits for $\left\langle \mu\right\rangle -1$ &$\left\langle \mu\right\rangle -1 \pm\sigma_{\mu}$\\
\hline
sz1  &$\left[ -2.13\cdot 10^{-4},3.46\cdot 10^{-4} \right] $ & $6.55\cdot 10^{-5}\pm1.09\cdot 10^{-4}$ & $\left[ 8.21\cdot 10^{-4}, 1.38\cdot 10^{-3} \right] $ &$1.10\cdot10^{-3}\pm1.10\cdot 10^{-4}$\\
ltb1 & $\left[ -2.07\cdot 10^{-4}, 3.50\cdot10^{-4}\right] $& $7.01\cdot 10^{-5}\pm1.08\cdot 10^{-4}$ & $\left[ 8.11\cdot 10^{-4},1.37\cdot 10^{-3} \right] $ & $1.09\cdot10^{-3}\pm 1.09\cdot10^{-4}$ \\
sz2 & $ \left[ -6.38\cdot 10^{-5}, 1.23\cdot 10^{-4} \right] $ & $2.94\cdot 10^{-5}\pm3.63\cdot 10^{-5}$ & $ \left[ 7.65\cdot 10^{-6}, 1.94\cdot 10^{-4} \right]   $ & $ 1.01\cdot 10^{-4}\pm 3.62\cdot 10^{-5} $ \\
ltb2 & $\left[ -5.66\cdot 10^{-5}, 1.27\cdot 10^{-4} \right] $ & $3.52\cdot 10^{-5}\pm3.58\cdot 10^{-5}$ & $ \left[7.65\cdot 10^{-6},1.94\cdot 10^{-4} \right]  $ & $9.12\cdot 10^{-5}\pm 3.57\cdot 10^{-5}$\\
\hline
\end{tabular}
\caption{Average values of $\mu$ and $\mu^{-1}$ with standard deviation and $99\%$ confidence intervals obtained with $10^{5}$ bootstrap samples.}
\label{table:mu}
\end{table*}

\newpage
\section{Conclusion}
Light propagation in four Swiss cheese models only differing in terms of the shapes of their individual structures was studied. In particular, the redshift-distance relation, shear and expansion rate were studied along light rays initialized equivalently in the four models. Despite the anisotropies of the Szekeres models, light rays initialized equivalently in the four models were all found to travel through equivalent portions of spacetime. The angular diameter distance along light rays in two of the models, models sz2 and ltb2, deviate significantly less from the background value than in the other two models, models sz1 and ltb1. This is consistent with the much more prominent density contrasts in the latter two models. While the angular diameter distance along equivalent light rays in models sz1 and ltb1 are similar, there is a noticeable difference between the angular diameter distances computed along the equivalent light rays of models sz2 and ltb2. This can presumably be attributed the particular anisotropies in the structures of models sz1 and sz2. Analogous similarities were found for the expansion rate and projected shear along the light rays. In addition, the integrals of the projected shear and of the fluctuations in the expansion rate along the light rays were found to cancel with each other to a high precision. This was also found in \cite{tardis} and it would be interesting to learn if it is a particular feature of Szekers/LTB models or if the feature persists for a larger group of spacetimes.
\newline\indent
The line-of-sight average of the angular diameter distance to the surface of last scattering, $\left\langle\delta D_{A,ls} \right\rangle $, was computed using $98304$ light rays for each model. For the two models with largest energy-density fluctuation amplitudes, $\left\langle\delta D_{A,ls} \right\rangle $ was found to be of order $10^{-4}$. For the two other models, it was two orders of magnitude smaller. For three of the four models, the shift was found to be negative. This is in agreement with perturbative results found in \cite{weinberg,angle_vs_ensemble_early,angle_vs_ensemble,Kaiser_Peacock}. Even in the single case of a positive average $\left\langle\delta D_{A,ls} \right\rangle $, the result cannot be claimed to be in discordance with perturbation theory since the result has a very low statistical significance.

\section{Acknowledgments}
The author would like to thank Niels Carl W. Hansen for identifying a function which was making the OpenMP parallelization of the original version of the code used for the work unmanageably slow. The author also thanks Steen Hannestad for valuable discussions on the work and Austin Peel and Mustapha Ishak for their correspondence. Lastly, the anonymous referees are thanked for their useful comments.
\newline\indent
The work presented here has been obtained with the use of computer resources from the Center for Scientific Computing Aarhus.


\begin{thebibliography}{99}
\bibitem{Linder1} Eric V. Linder: Averaging Inhomogeneous Universes: Volume, Angle, Line of Sight, arXiv:astro-ph/9801122v2 
\bibitem{syksy_av1} Syksy Rasanen: Light propagation in statistically homogeneous and isotropic dust universes, JCAP 0902:011,2009, arXiv:0812.2872v2 [astro-ph]
\bibitem{syksy_av2} Syksy Rasanen: Light propagation in statistically homogeneous and isotropic universes with general matter content,   JCAP 1003:018,2010 , arXiv:0912.3370v2 [astro-ph.CO]

\bibitem{bc1} Thomas Buchert: On average properties of inhomogeneous fluids in general relativity I: dust cosmologies,  Gen.Rel.Grav. 32 (2000) 105-125 , arXiv:gr-qc/9906015v2
\bibitem{bc2} Thomas Buchert: Toward physical cosmology: focus on inhomogeneous geometry and its non-perturbative effects, Class.Quant.Grav.28:164007,2011, arXiv:1103.2016v2 [gr-qc]
\bibitem{bc3} Thomas Buchert, Syksy Rasanen: Backreaction in late-time cosmology, Annual Review of Nuclear and Particle Science 62 (2012) 57-79, arXiv:1112.5335v2 [astro-ph.CO]
\bibitem{bc4} Chris Clarkson, George Ellis, Julien Larena, Obinna Umeh: Does the growth of structure affect our dynamical models of the universe? The averaging, backreaction and fitting problems in cosmology, Rept.Prog.Phys. 74 (2011) 112901, arXiv:1109.2314v1 [astro-ph.CO]

\bibitem{misinterp} Chris Clarkson, George Ellis, Andreas Faltenbacher, Roy Maartens, Obinna Umeh, Jean-Philippe Uzan: (Mis-)Interpreting supernovae observations in a lumpy universe,  Monthly Notices of the Royal Astronomical Society, Volume 426, Issue 2, pp. 1121-1136 (2012) , arXiv:1109.2484v3 [astro-ph.CO]
\bibitem{Linder_angle} Eric V. Linder: Transition from Clumpy to Smooth Angular Diameter Distances, Astrophys. J. 497, 28 (1998), arXiv:astro-ph/9707349v2



\bibitem{bolejko_DA} Krzysztof Bolejko: The effect of inhomogeneities on the distance to the last scattering surface and the accuracy of the CMB analysis, JCAP 02(2011)025, arXiv:1101.3338v1 [astro-ph.CO]




\bibitem{do_we_care} Camille Bonvin, Chris Clarkson, Ruth Durrer, Roy Maartens, Obinna Umeh: Do we care about the distance to the CMB? Clarifying the impact of second-order lensing, JCAP 1506 (2015) 06, 050, arXiv:1503.07831v3 [astro-ph.CO]
\bibitem{Kaiser_Peacock} Nick Kaiser, John A. Peacock: On the Bias of the Distance-Redshift Relation from Gravitational Lensing, Mon Not R Astron Soc (2016) 455 (4): 4518-4547,  arXiv:1503.08506v1 [astro-ph.CO]
\bibitem{Nbody_average} Joachim Wambsganss, Renyue Cen, Guohong Xu, Jeremiah P. Ostriker: Effects of Weak Gravitational Lensing from Large-Scale Structure on the Determination of $q_0$, Astrophys. J. 475, L81 (1997), arXiv:astro-ph/9607084v1
\bibitem{Nbody_LTB_average} Krzysztof Bolejko, Pedro G. Ferreira: Ricci focusing, shearing, and the expansion rate in an almost homogeneous Universe, JCAP05(2012)003, arXiv:1204.0909v2 [astro-ph.CO]
\bibitem{syksyCMB} Mikko Lavinto, Syksy Rasanen: CMB seen through random Swiss Cheese, JCAP10(2015)057, arXiv:1507.06590v3 [astro-ph.CO]
\bibitem{randomize} R. Ali Vanderveld, Eanna E. Flanagan, Ira Wasserman: Luminosity distance in "Swiss cheese" cosmology with randomized voids: I. Single void size, Phys.Rev.D78:083511,2008, arXiv:0808.1080v2 [astro-ph]
\bibitem{Ishak} Austin Peel, M. A. Troxel, Mustapha Ishak: Effect of inhomogeneities on high precision measurements of cosmological distances, Phys. Rev. D 90, 123536 (2014), arXiv:1408.4390v2 [astro-ph.CO]
\newline
Erratum: Phys. Rev. D 92, 029901 (2015)
\bibitem{tetris} Nikolaos Brouzakis, Nikolaos Tetradis, Eleftheria Tzavara: Light Propagation and Large-Scale Inhomogeneities, JCAP0804:008,2008, arXiv:astro-ph/0703586v4
\bibitem{tetris2} N. Brouzakis, N. Tetradis, E. Tzavara: The Effect of Large-Scale Inhomogeneities on the Luminosity Distance, JCAP 0702:013,2007, arXiv:astro-ph/0612179v2 
\bibitem{ltb_effect_hubble} Timothy Clifton, Joe Zuntz: Hubble Diagram Dispersion From Large-Scale Structure, Mon. Not. R. Astron. Soc. 400 (2009) 2185, arXiv:0902.0726v2 [astro-ph.CO]

\bibitem{LTB1} G. Lemaitre: L'Universe en expansion, Annales de la Societe Scientifique de Bruxelles  A53, 51 (1933)
\newline
English translation: General Relativity and Gravitation, Vol. 29, No. 5, 1997
\bibitem{LTB2} R. C. Tolman: Effect of Inhomogeneity on Cosmological Models, Proc. Natl. Acad. Sci. USA 20, 169-176 (1934)
\bibitem{LTB3} H. Bondi: Spherically Symmetrical Models in General Relativity, Month. Not. Roy. Astr. Soc. 107,410 (1947)


\bibitem{bias2} Valerio Marra, Edward W. Kolb, Sabino Matarrese, Antonio Riotto: On cosmological observables in a swiss-cheese universe, Phys.Rev.D76:123004,2007, arXiv:0708.3622v3 [astro-ph]
\bibitem{bias1} Valerio Marra, Edward W. Kolb, Sabino Matarrese: Light-cone averages in a swiss-cheese universe, Phys.Rev.D77:023003,2008, arXiv:0710.5505v2 [astro-ph] 


\bibitem{tardis} Mikko Lavinto, Syksy Rasanen, Sebastian J. Szybka: Average expansion rate and light propagation in a cosmological Tardis spacetime, JCAP12(2013)051, arXiv:1308.6731v2 [astro-ph.CO]
	

\bibitem{dig_selv} S. M. Koksbang, S. Hannestad: Methods for studying the accuracy of light propagation in N-body simulations, Phys. Rev. D 91, 043508 (2015), arXiv:1501.01413v2 [astro-ph.CO]
\bibitem{dig_selv2} S. M. Koksbang and S. Hannestad: Studying the precision of ray tracing techniques with Szekeres models, Phys. Rev. D 92, 023532 (2015), arXiv:1506.09127 [astro-ph.CO]
\newline Erratum: S. M. Koksbang and S. Hannestad, Phys. Rev. D 92, 069904 (2015)
\bibitem{LTB_in_NG} Karel Van Acoleyen: LTB solutions in Newtonian gauge: from strong to weak fields, JCAP0810:028,2008, arXiv:0808.3554v2 [gr-qc]
\bibitem{LTB_in_NG2} Aseem Paranjape, T. P. Singh: Structure Formation, Backreaction and Weak Gravitational Fields, JCAP 0803:023,2008, arXiv:0801.1546v3 [astro-ph]
\bibitem{LTB_in_NG3} T. Biswas and A. Notari: "Swiss-Cheese" Inhomogeneous Cosmology and the Dark Energy Problem, JCAP 0806:021,2008, arXiv:astro-ph/0702555v1 
\bibitem{bright_side} Krzysztof Bolejko, Chris Clarkson, Roy Maartens, David Bacon, Nikolai Meures, Emma Beynon: Anti-lensing: the bright side of voids, Phys. Rev. Lett. 110, 021302 (2013), arXiv:1209.3142v3 [astro-ph.CO]

\bibitem{Mexico} Roberto A. Sussman, Juan Carlos Hidalgo, Ismael Delgado Gaspar, Gabriel German: Non-Spherical Szekeres models in the language of Cosmological Perturbations, arXiv:1701.00819v1 [gr-qc]

\bibitem{bolejko_struct1} Krzysztof Bolejko: Structure formation in the quasispherical Szekeres model, Phys.Rev.D73:123508,2006, arXiv:astro-ph/0604490v2 
\bibitem{bolejko_struct2} Krzysztof Bolejko: Evolution of cosmic structures in different environments in the quasispherical Szekeres model, Phys.Rev. D75 (2007) 043508, arXiv:astro-ph/0610292v2
\bibitem{ishak_struct_vsLTB} M. A. Troxel, Austin Peel, Mustapha Ishak: Effects of anisotropy on gravitational infall in galaxy clusters using an exact general relativistic model, JCAP 1312:048, 2013, arXiv:1311.5651v2 [astro-ph.CO]
\bibitem{ishak_struct_vsLin1} Austin Peel, Mustapha Ishak, M. A. Troxel: Large-scale growth evolution in the Szekeres inhomogeneous cosmological models with comparison to growth data, Phys.Rev.D86,123508,2012, arXiv:1212.2298v1 [astro-ph.CO] 
\bibitem{ishak_struct_vsLin2} Mustapha Ishak, Austin Peel: The growth of structure in the Szekeres inhomogeneous cosmological models and the matter-dominated era, Phys. Rev. D85, 083502, 2012, arXiv:1104.2590v3 [astro-ph.CO]

\bibitem{angle_vs_ensemble} Camille Bonvin, Chris Clarkson, Ruth Durrer, Roy Maartens, Obinna Umeh: Cosmological ensemble and directional averages of observables, JCAP 1507 (2015) 07, 040, arXiv:1504.01676v2 [astro-ph.CO]
\bibitem{angle_vs_ensemble_early} T.W.B. Kibble, Richard Lieu: Average magnification effect of clumping of matter, Astrophys.J. 632 (2005) 718-726, arXiv:astro-ph/0412275v2 
\bibitem{weinberg} S. Weinberg: Apparent luminosities in a locally inhomogeneous universe, Astrophysical Journal, vol. 208:L1-L3 (1976)


\bibitem{Szekeres} P. Szekeres: A class of inhomogeneous cosmological models , Commun. Math. Phys. 41, 55 (1975)
\bibitem{Szekeres2} P. Szekeres: Quasispherical gravitational collapse,  Phys Rev D 12, 2941 (1975)
\bibitem{Szafron} D. A. Szafron: Inhomogeneous cosmologies: New exact solutions and their evolution, Journal of Mathematical Physics 18, 1673 (1977); doi: 10.1063/1.523468
\bibitem{killing} W. B. Bonnor, A. H. Sulaiman and N. Tomimura: Szekeres's space-times have no killing vectors, Gen. Relativ. Gravit. 8, 549 (1977)

\bibitem{killing_new} Ira Georg, Charles Hellaby: Symmetry and Equivalence in Szekeres Models, arXiv:1702.05347v1 [gr-qc] (preview)
 
\bibitem{voids_2001} M.Plionis, S.Basilakos: The Size and Shape of Local Voids, Mon.Not.Roy.Astron.Soc.330:399,2002, arXiv:astro-ph/0106491v2
\bibitem{Voids_2002} Fiona Hoyle, Michael S. Vogeley: Voids in the PSCz Survey and the Updated Zwicky Catalog, Astrophys.J.566:641-651,2002, arXiv:astro-ph/0109357v2 
\bibitem{voids_2003} Fiona Hoyle, Michael S. Vogeley: Voids in the 2dF Galaxy Redshift Survey, Astrophys.J.607:751-764,2004, arXiv:astro-ph/0312533v1

\bibitem{growing_modes} J. P. Zibin: Scalar Perturbations on Lemaitre-Tolman-Bondi Spacetimes, Phys.Rev.D78:043504,2008, arXiv:0804.1787v2 [astro-ph]
\bibitem{Acoleyen} Karel Van Acoleyen: Lemaitre–Tolman–Bondi solutions in Newtonian gauge: from strong to weak fields, JCAP0810:028,2008, arXiv:0808.3554v2 [gr-qc]

\bibitem{Jodrey} W. S. Jodrey and E. M. Tory: Computer simulation of close random packing of equal spheres, Phys. Rev. A 32, 2347 (1985), https://doi.org/10.1103/PhysRevA.32.2347
\newline
Erratum: Phys. Rev. A 34, 675 (1986)
\bibitem{Jodrey2} W. S. Jodrey and E. M. Tory: Computer simulation of isotropic, homogeneous, dense random packing of equal spheres. Powder Tech., 30:111–118, 1981, http://dx.doi.org/10.1016/0032-5910(81)80003-4
\bibitem{RCP_experimental} G. D. Scott and D. M. Kilgour: The density of random close packing of spheres, BRIT. J. APPL. PHYS. (J. PHYS. D), 1969, SER. 2, VOL. 2.

\bibitem{voids_77} Rien van de Weygaert: Voids and the Cosmic Web: cosmic depressions \& spatial complexity, 10.1017/S1743921316010504, arXiv:1611.01222v2 [astro-ph.CO]


\bibitem{poly1} Yixiang Gan, Marc Kamlah, Jorg Reimann: Computer simulation of packing structure in pebble beds, Fusion Engineering and Design 85 (2010) 1782–1787, http://dx.doi.org/10.1016/j.fusengdes.2010.05.042
\bibitem{poly2} Anuraag R. Kansal, Salvatore Torquato, and Frank H. Stillinger: Computer generation of dense polydisperse sphere packings,The Journal of Chemical Physics 117, 8212 (2002), http://dx.doi.org/10.1063/1.1511510


\bibitem{Ishak2} M. A. Troxel, Mustapha Ishak, Austin Peel: The effects of structure anisotropy on lensing observables in an exact general relativistic setting for precision cosmology, JCAP 1403:040, 2014, arXiv:1311.5936v2 [astro-ph.CO] 
 
\bibitem{god_bog} G. F. R. Ellis, R. Maartens, M. A. H. MacCallum: Relativistic cosmology, Cambridge university press 2012
\bibitem{lensing_bog} P. Schneider, J. Ehlers, and E. E. Falco: Gravitational Lenses (study edition) (Springer-Verlag, Berlin, 1999)
\bibitem{sachs} R. Sachs: Gravitational Waves in General Relativity. VI. The Outgoing Radiation Condition, November 1961 Proc. R.
Soc. Lond. A 1961 264, doi: 10.1098/rspa.1961.0202
\bibitem{arbitrary_spacetime} Stella Seitz, Peter Schneider, Jurgen Ehlers: Light propagation in arbitrary spacetimes and the gravitational lens approximation, Class.Quant.Grav.11:2345-2374,1994 , astro-ph/9403056
\bibitem{kappa_explained} Camille Bonvin, Sambatra Andrianomena, David Bacon, Chris Clarkson, Roy Maartens, Teboho Moloi, Philip Bull: Dipolar modulation in the size of galaxies: The effect of Doppler magnification, arXiv:1610.05946v1 [astro-ph.CO]
\bibitem{bonvin} Camille Bonvin: Effect of Peculiar Motion in Weak Lensing, Phys.Rev.D78:123530,2008, arXiv:0810.0180v3 [astro-ph]

\bibitem{cosmo_doppler} David J. Bacon, Sambatra Andrianomena, Chris Clarkson, Krzysztof Bolejko, Roy Maartens: Cosmology with Doppler Lensing, MNRAS 443 (2014) 1900-1915, arXiv:1401.3694v2 [astro-ph.CO] 

\bibitem{Ehlers} Jurgen Ehlers: Beitrage zur relativistischen Mechanik kontinuerlicher Medien, Proceedings of the Mathematical-Natural Science Section of the Mainz Academy of Science and Literature, Nr. 11, 1961 (pp. 792-837)
\newline
English translation by G. F. R. Ellis and P. K. S. Dunsby: Contributions to the Relativistic Mechanics of Continuous Media, General Relativity and Gravitation, Vol. 25, No. 12, 1993


\bibitem{bootstrap} B. Efron: Bootstrap methods: another look at the jackknife, The annals of statistics, 1979, Vol. 7, No. 1, 1-26






\end{thebibliography}
\end{document}